\documentclass[notitlepage,amsmath,preprintnumbers,superscriptaddress,nofootinbib,aps,11pt]{revtex4-1} 

\pdfoutput=1

\usepackage{amsmath,url}
\usepackage{amssymb,amsfonts}
\usepackage{rotating}
\usepackage{color}
\usepackage{hhline}

\usepackage{xcolor}
\usepackage{multirow}

\usepackage{array}

\usepackage{booktabs}

\definecolor{Gray}{gray}{0.85}
\definecolor{LightGreen}{rgb}{0.88, 1, 0.88}
\definecolor{LightCyan}{rgb}{0.88,1,1}
\definecolor{LightRed}{rgb}{1, 0.85, 0.85}
\definecolor{LightYellow}{rgb}{1, 1, 0.85}
\definecolor{LightBlue}{rgb}{0.87, 0.94, 1}
\definecolor{white}{gray}{1}
\definecolor{black}{gray}{0}
\newcolumntype{G}{>{\columncolor{LightGray}}c}
\newcommand{\cchi}{\cellcolor{LightBlue}}

\newcommand{\colb}{\rowcolor{LightBlue}}

\newcommand{\coln}{\cellcolor{LightRed}}
\newcommand{\coly}{\cellcolor{LightGreen}}

\usepackage{array,mathtools,amssymb,booktabs}
\newcolumntype{C}{>{$}c<{$}}
\AtBeginDocument{
\heavyrulewidth=.16em
\lightrulewidth=.1em
\cmidrulewidth=.03em
\belowrulesep=.4ex
\belowbottomsep=0pt
\aboverulesep=.4ex
\abovetopsep=0pt
\cmidrulesep=\doublerulesep
\cmidrulekern=.5em
\defaultaddspace=.5em
}
\usepackage{framed}
\usepackage{bm}
\usepackage{amsmath}
\usepackage{graphicx}
\usepackage{amssymb}
\usepackage{epstopdf}
\usepackage{hyperref}
\usepackage{subfigure}
\usepackage{epstopdf}
\usepackage{verbatim} 
\usepackage{array}
\usepackage{booktabs}
\usepackage{color}

\def\beq{\begin{equation}}
\def\eeq{\end{equation}}

\def\bea{\arraycolsep .1em \begin{eqnarray}}
\def\eea{\end{eqnarray}}
\def\Tr{{\rm Tr}}
\newcommand{\step}{\vspace{.5em}}

\def\dy{D}

\def\eq#1{(\ref{#1})}

\def\s0#1#2{\mbox{\small{$ \frac{#1}{#2} $}}}
\def\0#1#2{\frac{#1}{#2}}

\def\grgl{\:\hbox to -0.2pt{\lower2.5pt\hbox{$\sim$}\hss}{\raise3pt\hbox{$>$}}\:}
\def\klgl{\:\hbox to -0.2pt{\lower2.5pt\hbox{$\sim$}\hss}{\raise3pt\hbox{$<$}}\:}

\newcommand \be {\begin{equation}}
\newcommand \ee {\end{equation}}
\newcommand \bed {\begin{displaymath}}
\newcommand \eed {\end{displaymath}}

\newcommand{\bit}{\begin{itemize}}
\newcommand{\eit}{\end{itemize}}

\usepackage{colortbl}

\definecolor{Gray}{gray}{0.85}
\definecolor{LightGray}{gray}{0.93}
\definecolor{LightGreen}{rgb}{0.88, 1, 0.88}
\definecolor{LightCyan}{rgb}{0.88,1,1}
\definecolor{LightRed}{rgb}{1, 0.85, 0.85}
\definecolor{LightRed}{rgb}{1, 0.88, 0.88}
\definecolor{LightYellow}{rgb}{1, 1, 0.85}
\definecolor{LightBlue}{rgb}{0.87, 0.94, 1}
\definecolor{white}{gray}{1}

\newcommand{\gp}{\mathcal{G}}
\newcommand{\minrat}{\chi}

\makeatletter
    \def\CT@@do@color{%
      \global\let\CT@do@color\relax
            \@tempdima\wd\z@
            \advance\@tempdima\@tempdimb
            \advance\@tempdima\@tempdimc
    \advance\@tempdimb\tabcolsep
    \advance\@tempdimc\tabcolsep
    \advance\@tempdima2\tabcolsep
            \kern-\@tempdimb
            \leaders\vrule
                    \hskip\@tempdima\@plus  1fill
            \kern-\@tempdimc
            \hskip-\wd\z@ \@plus -1fill }
            
\begin{document}
${}$\vskip1cm

\title{Theorems for Asymptotic Safety  of Gauge Theories}

\author{Andrew~D.~Bond}
\email{a.bond@sussex.ac.uk}

\author{Daniel F.~Litim}
\email{d.litim@sussex.ac.uk}

\affiliation{\mbox{Department of Physics and Astronomy, U Sussex, Brighton, BN1 9QH, U.K.}}

\begin{abstract}
We classify the weakly interacting fixed points of general gauge theories coupled to matter and explain how the competition between gauge and matter fluctuations gives rise to a rich spectrum of high- and low-energy fixed points. The pivotal role played by  Yukawa couplings is emphasized. Necessary and sufficient conditions for asymptotic safety of gauge theories are also derived, in conjunction with strict no go theorems. Implications for phase diagrams of gauge theories and physics beyond the Standard Model are indicated.
\end{abstract}

\maketitle

{\bf 1.} 
Fixed points of the renormalisation group play an important role in quantum field theory and particle physics \cite{Wilson:1971bg,Wilson:1971dh}.  Low-energy fixed points characterise continuous phase transitions and the dynamical breaking of symmetry.  High-energy fixed points are central for the fundamental definition of quantum field theory. Important examples are provided by asymptotic freedom of non-abelian gauge theories \cite{Gross:1973id,Politzer:1973fx} where the high-energy fixed point is non-interacting. 
Gauge theories with complete asymptotic freedom, meaning asymptotic freedom for all of its couplings, are of particular interest in the search for extensions of the Standard Model \cite{Giudice:2014tma}. Asymptotically free gauge theories can also display weakly coupled infrared (IR) fixed points \cite{Caswell:1974gg,Banks:1981nn}. More recently, 
it was discovered that gauge theories can   develop   interacting ultraviolet (UV) fixed points  \cite{Litim:2014uca}, a scenario known as asymptotic safety. This intriguing new phenomenon, originally conjectured  in the context of quantum gravity  \cite{Weinberg:1980gg}, offers the prospect for consistent UV completions of particle physics beyond the  paradigm of asymptotic freedom \cite{Litim:2011cp}.  \step

In this Letter we classify all weakly interacting  fixed points of general gauge theories coupled to  matter in four space-time dimensions starting   from first principles. Our motivation for doing so is twofold: Firstly, we want to understand in general terms whether and how the competition between gauge and matter  field  fluctuations gives rise to quantum scale invariance. 
We expect that  insights into conformal windows of gauge theories will offer  new directions for particle  physics above the electroweak energy scale. Secondly, we are particularly interested in the dynamical origin for asymptotic safety in gauge theories and conditions under which it may arise. We also hope that insights into the inner working  of asymptotic safety at weak coupling  will offer clues for mechanisms of asymptotic safety  at strong coupling  \cite{Falls:2013bv,Falls:2014tra}.
\step

We  pursue these questions in perturbation theory starting with pure gauge interactions and gradually adding  in more gauge and matter couplings. We will find a rich spectrum of interacting high- and low-energy fixed points including necessary and sufficient conditions for their existence. Furthermore, we  highlight the  central importance of Yukawa couplings to balance gauge  against matter  fluctuations.  We thereby also  establish that the presence of  scalar fields such as the Higgs are strict necessary conditions for asymptotic safety at weak coupling. Further key ingrediences for our results are bounds on quadratic Casimirs which are derived  for general Lie algebras, together with  structural aspects of  perturbation theory which are detailed as we proceed.

{\bf 2.} 
We begin our investigation of weakly coupled fixed points 
by considering (non-)abelian vector gauge theories with a simple gauge group $\gp$ and gauge coupling $g$, interacting with  spin-$\s012$ fermions or scalars or both. Throughout we scale loop factors into the definition of couplings  and introduce $\alpha= {g^2}/{(4\pi)^2}$. The renormalisation group running of the gauge coupling up to two loop order in perturbation theory reads
\beq\label{eq:GaugeBeta}
	\beta= - B\,\alpha^2 + C\, \alpha^3+{\cal O}(\alpha^4)\,,
\eeq
where $\beta\equiv{d\alpha}/{d(\ln \mu)}$, and $\mu$ denoting the RG momentum scale. The  one and two loop coefficients in \eqref{eq:GaugeBeta} are known for arbitrary field content and given in \cite{GellMann:1954fq,Bogolyubov:1980nc,Gross:1973id, Politzer:1973fx,Gross:1973ju} and \cite{Jones:1974mm, Caswell:1974gg,Jones:1974pg}, respectively. In terms of the  Dynkin index $S_2^R$ and  the quadratic Casimir $C_2^R$ of quantum fields in some irreducible representation (irrep) $R$ of the gauge group, they can be written as\footnote{Throughout, we treat fermions as Weyl and scalars as real.} 
\begin{align}
	B &= \0{2}{3}\left(11 C_2^\gp - 2 S_2^F - \0{1}{2}S_2^S\right)\,,\label{eq:OneLoop}\\
	C &= 2\left[\left( \0{10}{3}C_2^\gp+2C_2^F\right)S_2^F 
		+ \left(\0{1}{3}C_2^\gp+2 C_2^S\right)S_2^S - \0{34}{3}(C_2^\gp)^2\right]\,.\label{eq:TwoLoop}
\end{align}
The terms involving  $C_2^\gp$ -- the quadratic Casimir in the adjoint representation of the gauge group -- arise due to the fluctuations of the gauge fields. The fluctuations of charged fermionic (F) or scalar (S) matter fields, if present, contribute to \eq{eq:GaugeBeta} via the terms proportional to the Dynkin index of their representation.\step

Gauge theories with \eq{eq:GaugeBeta} will always display the free Gaussian fixed point $\alpha_*=0$. If $B>0$ this is the well-known ultraviolet (UV) fixed point of asymptotic freedom \cite{Gross:1973id,Politzer:1973fx} such as in QCD. For $B<0$, instead, the theory becomes free in the infrared (IR) such as in QED. In addition, \eq{eq:GaugeBeta} can also display an interacting fixed point
\beq\label{BC}
\alpha_*=\frac{B}{C}
\eeq
which is perturbative if $\alpha_*\ll 1$ and physically acceptable provided that $B\cdot C>0$. For $B\cdot C<0$ the would-be fixed point reads $\alpha_*<0$ and resides in an unphysical regime where the theory is sick non-perturbatively \cite{Dyson:1952tj}.  
Also, if $B<0$ ($B>0$), \eq{BC}  corresponds to an interacting UV (IR) fixed point. We conclude that the availability and nature of interacting fixed points is encoded in the signs and magnitude of \eq{eq:OneLoop} and \eq{eq:TwoLoop}.
From the explicit expressions, we observe that 
the pure gauge contributions to both the one and two~loop terms 
are either negative (non-abelian)  or vanishing (abelian).  Conversely, terms originating from fermionic or scalar matter contribute positively. This means that with a sufficiently small amount of matter (including none), the gauge boson contributions dominate and we have $B > 0, C < 0$. On the other hand, for a sufficiently large amount of matter, the matter contributions dominate and we end up with $B \le 0, C > 0$. 
The latter is trivially the case  for abelian gauge groups whose quadratic Casimir vanishes identically, $C_2^{U(1)} = 0$. Weakly interacting fixed points are absent in either of these cases.
\step

The question of what may happen when the pure gauge and matter contributions are of similar size is not immediately obvious. It has long been known that it is possible for theories to have $B, C > 0$, which are therefore asymptotically free and which, if $B \ll C$, can lead to a perturbative infrared 
Banks-Zaks fixed point \cite{Caswell:1974gg,Banks:1981nn}. 
However, no examples have been found for which $ B, C < 0$ and where the analagous fixed point would be ultraviolet. To see if such a scenario is possible in principle, we must examine the relative effects of matter on the one- and two-loop contributions. To that end, we resolve \eq{eq:OneLoop} for the adjoint Casimir and insert the result into the last term of \eq{eq:TwoLoop} to find 
\begin{align}\label{eq:GenericCBound}
	C = \frac{2}{11}  \Big[2 S_2^F \left(11 {C_2^F}+7C_2^\gp 
	\right) + 2 S_2^S \left(11 {C_2^S}-C_2^\gp\right) - 17 B\,C_2^\gp\Big]\,.
\end{align} 
We make the following obervations. The first term in \eq{eq:GenericCBound} due to the fermions is manifestly positive-definite. The last term in \eq{eq:GenericCBound} is positive-definite provided that $B< 0$. Hence, as has  been noted  by Caswell  \cite{Caswell:1974gg}, fermionic matter alone cannot generate an asymptotically safe UV fixed point in perturbation theory. 
The middle term however, due to charged scalars, is not manifestly positive definite and it cannot be decided {\it prima facie} whether or not it may generate an interacting UV fixed point with $B<0$ and $C<0$. 
\step

{\bf 3.} 
In order to progress with the analysis of \eq{eq:GenericCBound}, we must find expressions for the smallest quadratic Casimir for any simple Lie algebra ${\cal G}$.
Irreducible representations of simple Lie algebras are conveniently characterised by their highest weight $\Lambda$, which for a rank-$n$ Lie algebra is an $n$-dimensional vector of non-negative integers, not all of which are zero.\footnote{We are not interested in trivial representations given that uncharged fields cannot contribute to \eq{eq:GaugeBeta}.}
 This is due to the theorem of highest weight, which states that inequivalent irreps are in one-to-one correspondence with distinct highest weights. The Racah formula offers an explicit expression for the quadratic Casimir for any irrep $R$ with highest weight $\Lambda$. It is given by 
\begin{align}\label{eq:Racah}
	C_2(\Lambda) &= \0{1}{2}(\Lambda, \Lambda + 2\delta)\,,
\end{align}
where 
$(u,v)\equiv \sum_{ij}G_{ij}\,u^i\,v^j$ 
denotes the inner product of two highest weights, 
with $u=\sum_{i=1}^n u^i\,\Lambda_i$. The weight metrics $G\equiv (G_{ij})$ 
are known explicitly for any Lie algebra ${\cal G}$. Note that $(u,v)> 0$ for any two weights.
The $n$-component vector $\delta$  in \eq{eq:Racah} denotes half the sum of the positive roots and reads	$\delta= (1,1,\dots ,1)$
in the Dynkin basis (which we use exclusively).  The normalisation factor $\frac{1}{2}$ in \eq{eq:Racah} is conventional.\footnote{In general, the quadratic Casimir is only defined up to a multiplicative constant for a given Lie algebra, and thus we are free to choose the overall normalisation.} \step

For any Lie algebra,  the highest weight of irreps with the smallest quadratic Casimir must be one of the fundamental weights  $\Lambda_k$ (with $k \in \{1,\dots,n\}$), whose components are defined as
\begin{align}\label{min}	(\Lambda_k)^i &= \delta^i_k\,.
\end{align}
This can be understood as follows. 
Consider  two highest weights $\Lambda$ and $\lambda$, which may be used to construct a new irrep with highest weight $\Lambda+\lambda$. The bilinearity of the inner product \eq{eq:Racah} then implies that
\beq\label{sum}
C_2(\Lambda+\lambda)
> C_2(\Lambda)+C_2(\lambda)
> C_2(\Lambda)\,.
\eeq
It follows, trivially, that $C_2$ can be made arbitrarily large. To find the smallest $C_2$, however, \eq{sum}  states that we only need to consider irreps whose highest weights have a single non-vanishing component. Assuming $\Lambda$ to be one such weight and taking $\lambda=m\,\Lambda$ for some integer $m\ge 1$, \eq{sum} also states that we only need to consider highest weights where this single non-vanishing component takes the smallest non-vanishing value, which is unity. This establishes \eq{min}.
Inserting \eq{min} into \eq{eq:Racah}, and denoting by $G$ the weight metric of the gauge group $\gp$, we find the quadratic Casimir in terms of the fixed index $k$ as
\beq\label{C2k}
	C_2= 	 \frac{1}{2}G_{kk}+\sum_{i=1}^n G_{ki}\,.
	 \eeq 
It remains to  identify the minima of \eq{C2k} with respect to $k$ for the  
four classical and  the five exceptional Lie algebras  separately, following the Cartan classification, starting with the rank-$n$ classical Lie algebras $A_n, B_n, C_n$ and $D_n$  \cite{humphreys1972introduction}. For $n \geq 1,2,3$ and $4$ they correspond to the unique Lie algebras  ${\bf su}(n+1), {\bf so}(2n+1), {\bf sp}(n)$ and ${\bf so}(2n)$, respectively. Explicit expressions for the weight metrics are summarised in~\cite{Slansky:1981yr}. For our purposes 
we write them in closed form as  
\begin{align}
	(G^{A_n})_{ij} &= \min(i,j)
- \frac{ij}{n+1}
\,,\nonumber \\
	(G^{B_n})_{ij} &= \frac{1}{2}\left[\min(i,j)(2 - \delta_{in} - \delta_{jn} ) + \frac{n}{2}\delta_{in}\delta_{jn}\right]\,,\nonumber \\
	(G^{C_n})_{ij} &= \frac{1}{2}\min(i,j)\,,\nonumber \\
	(G^{D_n})_{ij} &=\frac{1}{2}\left[\min(i,j)\left(2 - \delta_{in} - \delta_{jn} -\delta_{i,n-1} - \delta_{j,n-1}\right) 
		+ \frac{n}{2}\left(\delta_{i,n-1}\delta_{j,n-1}+\delta_{in}\delta_{jn}\right) \right.\nonumber\\
		\label{metric}
	&\left.\qquad\qquad+ \frac{1}{2}(n-2)\left(\delta_{i,n-1}\delta_{j,n}+\delta_{i,n}\delta_{j,n-1}\right)\right]\,.
\end{align}
For illustration, we consider explicitly the case for $A_n$, where $G_{kk} = {k(n+1 - k)}/{(n+1)}$, which, combined with 
\begin{align}
	\sum_{i=1}^n G_{ki}&= \sum_{i=1}^k i 
				 +\sum_{i=k+1}^n k
				 -\frac{k}{n+1}\sum_{i=1}^n i\nonumber
				 = \frac{1}{2}k\left(n+1 - k\right)\,,
\end{align}
leads to the desired expression for $C_2(A_n)$ as stated in \eq{C2kall} below. Analoguous, if slightly more tedious, intermediate steps for the other cases lead to the result 
\begin{align}
	C_2(A_n)&=\frac{k}{2}\frac{(n+1 - k)(n+2)}{n+1}\,,\nonumber \\
	C_2(B_n) &= 
		\frac{1}{2}\left(k(2n+1-k) - \frac{1}{4}n(3+2n)\delta_{kn}\right)\,,\nonumber \\
	C_2(C_n) &=  \frac{k}{2}\left(n+1 - \frac{1}{2}k\right)\,,\nonumber \\
\label{C2kall}
C_2(D_n) &=  \frac{1}{2}\left(k(2n-k) 
			-\frac{n}{4}\left( 2n - 3 + 4k\right)(\delta_{k,n-1} +\delta_{kn})\right)\,,
\end{align}
with $k$ taking values between $1$ and $n$. To find the global minima of the expressions \eq{C2kall} with respect to $k$, we proceed as follows. For $A_n$ and $C_n$, the expressions are quadratic polynomials in $k$ with negative $k^2$ coefficient, implying that its minima are achieved  at the boundaries, meaning either $k=1$ or $k=n$, or both.  For $B_n$ and $D_n$, additionally, the expressions are discontinuous for certain intermediate values of  $k$ (owing to the $\delta_{k,n-1}$ and $\delta_{kn}$ factors). This implies that global minima may additionally be achieved for integer values of $k$ within the interval $(1,n)$. With this in mind, and after evaluating all possible cases, the final result for the smallest quadratic Casimir for the classical Lie algebras is found to be
\begin{align}
		{\min}\ C_2(A_n) &= \0{n}{2}\frac{n+2}{n+1}\,,\nonumber \\
		{\min}\ C_2(B_n) &=\left\{
     				\begin{array}{lr}
      				 	\frac{1}{8}n(2n+1) & {\rm for}\ \ n = 2,3\\
       					n    & {\rm for}\ \  n \geq 4\ \ \  \
     					\end{array}\,,
   			\right.\nonumber \\
	{\min}\ C_2(C_n) &= \frac n 2 + \frac 1 4\,,\nonumber \\
	\label{ABCDmin}
	{\min}\ C_2(D_n) &= n - \frac{1}{2}\,.
\end{align}
The five exceptional groups $E_{6,7,8}, F_4$, and $G_2$ have a fixed size, hence finding the smallest Casimir amounts to a simple minimisation. Using the appropriate expressions for the weight metrics \cite{Slansky:1981yr}, our results are summarised in Tab.~\ref{tab:AllGroupMins} where, for convenience, we  express \eq{ABCDmin} using the particle physics nomenclature for the gauge groups.\step

\begin{table}[t]
\begin{center}
\begin{tabular}{c |c c cc l}
 \rowcolor{LightGreen}
\toprule
&&&&\cchi&\\
&&&&\cchi&\\[-5mm]
\rowcolor{LightGreen}
\multirow{-2}{*}{\ symmetry\ } & 
\multirow{-2}{*}{range} & 
\multirow{-2}{*}{${\rm min}\ C_2$} &
\multirow{-2}{*}{\ \ $C_2({\rm adj})$\ \ }& 
\multirow{-2}{*}{\cchi$\bm{\minrat}$}& 
\multirow{-2}{*}{\ \ \ \ \ irrep with smallest $C_2$}
\\
\midrule
&&&&\cchi&\\[-5mm]
 \rowcolor{white} &&&&\cchi&\\[-3.5mm]
 \rowcolor{white}
$\bm{SU(N)}$& 
\footnotesize{$ N \geq 2$}& 
$\frac{N^2-1}{2N}$	&
$N$		&
\cchi\  $\bm{\frac{1}{2}\left(1-\frac{1}{N^2}\right)} \ $&
\ \ fundamental $\bm N$ and $\overline{\bm N}$
\\[1mm]
 \rowcolor{LightGray}
&&&&\cchi&\\[-3.mm]
 \rowcolor{LightGray}
& \ \footnotesize{$3 \leq N \leq 7$} \  
&\ $\frac{1}{16}N(N-1)$ \ &	
 $N - 2$	&\cchi
 $ \bm{\frac{N}{16}\frac{N-1}{N-2}}$& 
 \ \ fundamental spinors $\bf 2^{\lceil {\bm N}/{2} \rceil - 1}$ 
\\[.5mm]
 \rowcolor{white}
\cellcolor{LightGray}&&&&\cchi&\\[-3mm]
\rowcolor{white}
\cellcolor{LightGray}
& 
&
\cellcolor{white}
&
&
\cchi& 
\ \ fundamental vector $\bf 8_v$ and \\
\rowcolor{white}
\cellcolor{LightGray}
\multirow{-2}{*}{\cellcolor{LightGray}$\bm{SO(N)}$} &\multirow{-2}{*}{\footnotesize{$N = 8$}}
&\multirow{-2}{*}{$\frac{7}{2}$}
&\multirow{-2}{*}{$6$}
&\multirow{-2}{*}{\cchi$\bm{\frac{7}{12}}$} 
& \ \ fundamental  spinors $\bf  8_s$, $\bf 8_c$
\\
\rowcolor{LightGray}
\cellcolor{LightGray}&&&&\cchi&\\[-3mm]
\rowcolor{LightGray}
& 
\footnotesize{$N \geq 9$}&
$\frac{1}{2}(N-1)$&
$N - 2$&
\cchi$\bm{\frac{N-1}{2(N-2)}}$  
& \ \ fundamental $\bm N$ 
\\[.9mm] 
\rowcolor{white}
&&&&\cchi&\\[-3mm]
\rowcolor{white}
$\bm{Sp(N)}$& \footnotesize{$N \geq 1$}	& $\frac{1}{4}(2N + 1)$	& $N + 1$& \cchi$\bm{\frac{2N+1}{4(N+1)}}$ &\ \ fundamental $\bm{2N}$\\[.9mm]
\rowcolor{LightGray}
&&&&\cchi&\\[-3mm]
 \rowcolor{LightGray}
\multicolumn{1}{c|}{\cellcolor{LightGray}$\bm{E_8}$}&	& 30	& 30	&\cchi $\bm 1$& \ \ adjoint {\bf 248}	\\ [1ex] 
\rowcolor{white}
&&&&\cchi&\\[-3mm]
\rowcolor{white}	
\multicolumn{1}{c|}{$\bm{E_7}$}&		& $\frac{57}{4}$	& 18	&\cchi$\bm{\frac{19}{24}}$& \ \ fundamental {\bf 56}	\\ [1ex] 
\rowcolor{LightGray}&&&&\cchi&\\[-2mm]	
\rowcolor{LightGray}	
\multicolumn{1}{c|}{\cellcolor{LightGray}$\bm{E_6}$}	&&$\frac{26}{3}$	& 12	
			&\cchi$\bm{\frac{13}{18}}$&\ \  fundamental  {\bf 27} and $\overline{\bf 27}$	\\ [1ex] 
\rowcolor{white}
&&&&\cchi&\\[-3mm]	\rowcolor{white}\multicolumn{1}{c|}{$\bm{F_4}$}	&& 6	& 9	&\cchi$\bm{\frac{2}{3}}$&\ \  fundamental {\bf 26}	\\ [1ex] 
 \rowcolor{LightGray}
&&&&\cchi&\\[-3mm]\rowcolor{LightGray}
	\multicolumn{1}{c|}{\cellcolor{LightGray}$\bm{G_2}$}	&& 2	& 4	&\cchi$\bm{\frac{1}{2}}$&\ \  fundamental {\bf 7}
		\\[1ex]
		\bottomrule
\end{tabular}
\end{center}
 \caption{Summary of minimal Casimirs for the classical and exceptional Lie algebras along with the Casimir in the adjoint, their ratio $\minrat$, and the representations that attain the minimum. We notice that for $D_4$, corresponding to $SO(8)$, the Dynkin diagram has a three-fold symmetry leading to triality amongst the smallest Casimirs in the fundamental vector and spinor representations.} \label{tab:AllGroupMins}
\end{table}

A few comments are in order: $(i)$ For $A_n$ either boundary is minimal, corresponding to the fundamental and anti-fundamental representation. $(ii)$ For $B_n$ the Casimir is minimal for $k=n$ (the fundamental spinor representation) provided $n=2$ or $3$, and for $k=1$ (the fundamental vector representation) provided $n \geq 4$.  $(iii)$ For $C_n$ and $D_n$, the Casimir is minimal for $k = 1$ (the fundamental vector representation). $(iv)$  For $D_4$, three smallest Casimirs are achieved for $k = 1, 3$ and $4$. This degeneracy is due to the fact that the Dynkin diagram for $D_4$ possesses a three-fold symmetry, and thus there is a triality between the fundamental vector and the two inequivalent spinor representations.   $(v)$  For the exceptional groups, we find that the smallest Casimir is unique, except for $E_6$. $(vi)$   $E_8$ is the only group where the smallest Casimir is achieved for the adjoint representation (which is also one of the fundamental representations). $(vii)$  While the quadratic Casimir in general is a non-monotonic function of the dimensionality of the representation, our findings establish that the smallest Casimir always corresponds to those representations with the smallest dimension, which is always one of the fundamental representations.\step

\begin{figure}[t]
\begin{center}
\hskip-.6cm\includegraphics[scale=.8]{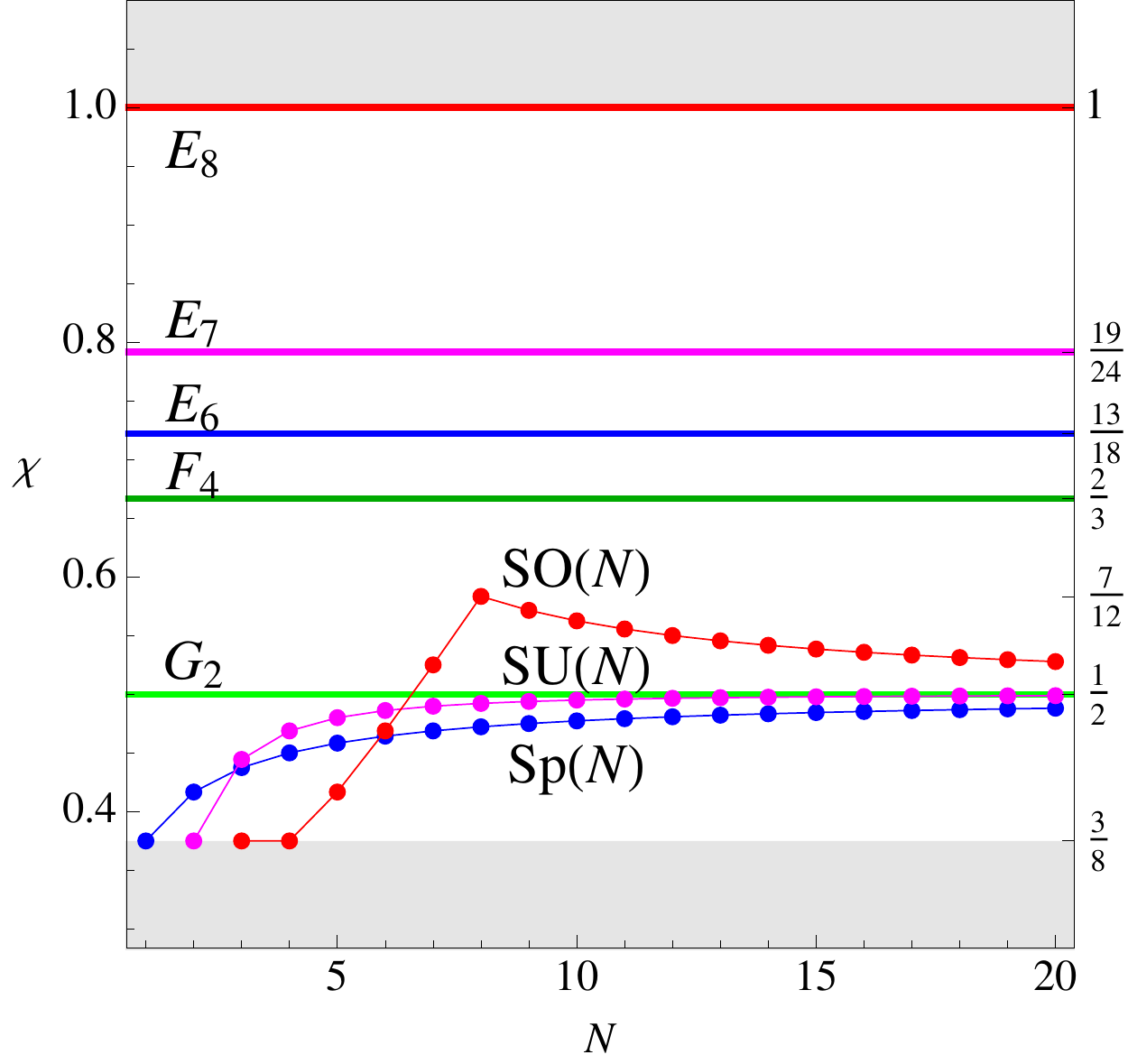}
\caption{Shown is the ratio $\chi$ \eq{chi} -- the smallest achievable quadratic Casimir in units of the Casimir in the adjoint -- for all simple Lie algebras.  The gray areas show the excluded domains. We observe that $\frac38\le \chi\le 1$. The lower bound 
is achieved for the fundamental two-dimensional representation of $SU(2) \simeq SO(3) \simeq Sp(1)$, and for the two inequivalent two-dimensional representation of $SO(4)$. For the exceptional groups the smallest Casimir grows with the rank of the group. The upper bound 
is achieved for $E_8$. In all cases, the smallest quadratic Casimir is achieved for the irreducible representation of smallest dimensionality. }\label{pCasimir} 
\end{center}
\end{figure}

Since the overall normalisation of quadratic Casimirs \eq{eq:Racah} can be chosen freely,  it is useful to consider the ratio  between the smallest quadratic Casimir and  the Casimir in the adjoint, 
\beq\label{chi}
\chi=\frac{{\rm min}\ C_2(R)}{C_2({\rm adj})}\,,
\eeq
which is independent of the normalisation. Fig.~\ref{pCasimir} shows our results for $\chi$ for all simple Lie algebras.   Evidently, $\chi$ is going to be bounded from above $\chi\le 1$ because the adjoint representation always exists. The upper boundary is achieved for the exceptional group $E_8$. Furthermore, $\chi$ is also bounded from below,
\beq\label{bound}
\0 38\le \chi \le 1\,.
\eeq
The lower bound is achieved  for the fundamental two-dimensional representation of $SU(2) \simeq SO(3) \simeq Sp(1)$, and for the two inequivalent two-dimensional representation of $SO(4)$.  We observe that $\chi$ is an increasing function with $N$ for $SU(N)$ and $Sp(N)$, interpolating between $\s038$ for small $N$ and $\s012$ in the infinte-$N$ limit. For $SO(N)$, we find that $\chi$ grows  from $\s038$  to its maximum $\s07{12}$ at $N=8$, from where it decays with increasing $N$ towards $\s012$ from above. From the exceptional groups, only $G_2$ has a $\chi$ value close to those of the classical groups. All other exceptional groups have larger values for $\chi$, which furthermore increases with the rank of the group.  \step

{\bf 4.} 
We are now in a position to develop the central results of this work, summarised in Tab.~\ref{tFinal} and Tab.~\ref{tFP}. 
We have observed   in \eq{eq:GenericCBound} that charged scalars potentially may turn the two loop coefficient $C$  negative even if $B\le 0$, provided that nontrivial scalar irreps are found with
$C^S_2 <\s0{1}{11} C_2^{\cal G}$.
However, the result \eq{chi}, \eq{bound} now firmly establishes that this is out of reach  for any simple Lie algebra, owing to $C^S_2 \ge \s0{3}{8} C_2^{\cal G}$.   Moreover, we find that the two loop coefficient obeys
\beq\label{Cbound}
C\ge C_2^\gp\left(\0{89}{22} S_2^F+\0{25}{22} S_2^S-\0{34}{11} B
\right)
\eeq
for any non-abelian gauge theory. Hence, while it is possible to have $B$ parametrically small such as in a Veneziano limit with suitably rescaled gauge coupling  \cite{Veneziano:1979ec}, the result \eq{Cbound} also shows that it is impossible to have both $B$ and $C$ parametrically small. Most importantly, we  conclude that for any gauge theory with a vanishing or positive one~loop coefficient for its gauge coupling's $\beta$~function, the two~loop coefficient is necessarily positive, 
\beq\label{nogo}
B\le 0\,\quad\Rightarrow\quad C>0\,,
\eeq
see \eq{eq:GaugeBeta}. It is  worth noting that  \eq{nogo} is not an equivalence: while $C<0$ arises exclusively only  if $B>0$, the case $C>0$ can arise irrespective of the sign of $B$
 \cite{Caswell:1974gg,Banks:1981nn}. Consequently, Banks-Zaks fixed points are invariably IR fixed points.
From the viewpoint of the asymptotic safety conjecture, our result \eq{nogo}  has the form of a no go theorem:  within perturbation theory, irrespective of the matter content and in the absence of non~gauge interactions, asymptotic safety cannot be realised for any four-dimensional  simple  non-abelian, or abelian, gauge theory.\footnote{Caswell has observed some time back that
{\it ``We do not expect to find a gauge theory of the above type} [meaning with \eq{eq:GaugeBeta}] {\it where $\beta$ starts out positive and goes negative near enough to the origin for the zero to be valid in perturbation theory.''} \cite{Caswell:1974gg}. Our result \eq{nogo} offers a general proof for Caswell's conjecture.}\step

The result \eq{nogo} straightforwardly generalises to  matter fields in generic reducible representations under the gauge symmetry. In this case it suffices to replace terms involving Dynkin indices and matter Casimirs in the one and two loop coefficients by
\beq
\label{Bsum}
S^R_2\rightarrow \sum_i S_2^{R_i}\,,\quad
S^R_2\, C^R_2\rightarrow \sum_i S_2^{R_i} \, C_2^{R_i}\,,
\eeq 
where the sums run over the decomposition into irreducible representations of the fermionic $(R=F)$ and scalar $(R=S)$ matter fields. Applying \eq{Bsum} to the two~loop coefficient  \eq{eq:GenericCBound}, 
we find that all fermionic contributions remain manifestly positive definite, and that each summand of the scalar contributions is positive definite owing to \eq{chi}, \eq{bound}.  We conclude that the no~go~theorem \eq{nogo} holds true for general matter representations, as summarised in Tab.~\ref{tFinal}~{\bf b)}.

{\bf 5.} 
Turning to more general gauge interactions, we consider gauge theories with product gauge groups ${\cal G}\equiv\otimes_{a=1}^n{\cal G}_a$ and multiple gauge couplings $\alpha_a$, each associated with a simple or abelian  factor $\gp_a$. We assume the presence of scalar and/or fermionic matter fields, some or all of which are charged under some or all of the gauge symmetries. In the absence of Yukawa interactions, the $\beta$~functions for the gauge couplings up to two loops in perturbation theory are of the form
\begin{align}\label{eq:MultiGaugeBeta}
	\beta_a &= \alpha_a^2\left(- B_a  + C_{ab} \,\alpha_b\right)+{\cal O}(\alpha^4)\,,
\end{align}
and $a,b=1,\cdots,n$. The coefficients $B_a$ and $C_{aa}$ (no sum) are the standard one and two~loop coefficients of the gauge coupling $\alpha_a$ as
given in \eqref{eq:OneLoop}, \eqref{eq:TwoLoop}. The new terms at two~loop level are the off-diagonal contributions $C_{ab}$ $(a\neq b)$ which parametrise the ${\cal O}(\alpha_b)$~contributions to the renormalisation group flow of couplings $\alpha_a$.
Nontrivial mixing between two gauge couplings arises through matter fields which are charged under both of these. The mixing terms can then be written as \cite{Nanopoulos:1978hh,Jones:1981we}
\beq\label{Cab}
	C_{ab}	= 4\left(C_2^{F_b} S_2^{F_a} + C_2^{S_b} S_2^{S_a} \right) \quad (a\neq b)\,.	
\eeq
The subscripts $a,b$ on
the Casimir or Dynkin index of the matter fields indicate the subgroup of $\gp$. From \eq{Cab} it follows that the mixing terms are manifestly non-negative
($C_{ab}\geq 0$)  for any semi~simple quantum gauge theory with or without abelian factors. The expression \eq{Cab} has a straightforward generalisation for reducible representations. Furthermore, if the theory contains more than one abelian factor, the off-diagonal contributions take a slightly different form in the presence of kinetic mixing \cite{delAguila:1988jz,Luo:2002iq}. In either of these cases, the mixing terms remain manifestly non-negative  ($C_{ab}\geq 0$, $a\neq b$). Together with  \eq{nogo} for all diagonal entries, we find that
\beq\label{nogoa}
B_a\le 0\,\quad\Rightarrow\quad C_{ab}\ge 0\quad{\rm for\ all\ \, }b\,,
\eeq
meaning that for  every infrared free gauge group factor ${\cal G}_a$, the corresponding column of the two~loop gauge contribution matrix $(C_{ab})$ is non-negative.\step

The result \eq{nogoa} has immediate implications for interacting fixed points of quantum field theories with \eq{eq:MultiGaugeBeta}, which, to leading order in perturbation theory, are given by all solutions of the linear equations 
\beq\label{FPab}
B_a=C_{ab}\,\alpha_b^* \,,\quad
{\rm subject\ to}\quad \alpha^*_b\ge 0\,.
\eeq
Assuming that  $B_a\le 0$ for at least one of the  subgroups ${\cal G}_a$, 
it follows from  \eq{nogoa}  that for \eq{FPab} to have a solution, at least one of the fixed points $\alpha_b^*$ must take negative values. However, we have already explained that such solutions are inconsistent \cite{Dyson:1952tj}, and
  conclude that the theory cannot have physically acceptable interacting fixed points within the perturbative regime as soon as any of the gauge factors is infrared free ($B_a\le 0$). In other words, the result \eq{nogoa} has the form of a no go theorem: asymptotic safety cannot be achieved for any semi-simple quantum gauge theory of the type \eq{eq:MultiGaugeBeta} with or without abelian factors and irrespective of the matter content. 
  \step
 
 Reversing the line of reasoning, our findings  also establish that physically-acceptable interacting fixed points in gauge theories with \eq{eq:MultiGaugeBeta} and without Yukawa interactions can only be achieved if all gauge group factors are asymptotically free $(B_a>0)$, which excludes $U(1)$ factors straightaway, see Tab.~\ref{tFP}~{\bf b)}. 
All weakly interacting fixed point solutions of \eq{FPab} are necessarily IR fixed points of the Banks-Zaks type inasmuch as they arise from balancing one and two loop gauge field flucutations. 
They also display a lesser number of relevant directions than the asymptotically free Gaussian UV fixed point meaning that  UV-IR connecting trajectories
exist which flow from the Gaussian down to any of the interacting fixed points. 
\step

Next, we investigate scalar and Yukawa-type matter couplings, and clarify whether these may help to generate weakly interacting fixed points. 
\step

{\bf 6.}   Scalar self-interactions  arise unavoidably in settings with charged scalars owing to the fluctuations of the gauge fields or in settings with uncharged scalars as long as these couple indirectly to the gauge fields through charged fermions and Yukawa interactions.
 Quartic scalar  self interactions or cubic ones in a phase with spontaneous symmetry breaking renormalise the gauge couplings starting at the three loop (four loop) level in perturbation theory,  provided the scalars are charged (uncharged) \cite{Curtright:1979mg}.\step
 
 In the light of \eq{nogo}, to help generate an interacting fixed point in the gauge sector once $B\le 0$, the scalar couplings would have to outweigh  the one~loop as well as the two~loop gauge contributions.   Even if the one~loop term vanishes identically ($B=0$), the result \eq{bound} together with \eq{eq:OneLoop}, \eq{eq:GenericCBound} and \eq{Cbound}
establishes that the two loop gauge coefficient is strictly positive $C(B=0)\ge C_{\rm min}$ and of order unity, with
\beq\label{Cmin}
C_{\rm min}/ (C^\gp)^2= 22\s014\,.
\eeq
The absolute minimum \eq{Cmin} is achieved for $Sp(1), SU(2), SO(3)$ and $SO(4)$ gauge symmetries.  The bound becomes slightly stronger with increasing $N$, reaching $C_{\rm min}/ (C^\gp)^2=25$
for the classical Lie groups in the infinite $N$ limit. For the exceptional groups $G_2, F_4, E_6,E_7$ and $E_8$ we find the increasingly stronger bounds $C_{\rm min}/ (C^\gp)^2=25,
50\s023,55\s059, 61\s023$
and $80$, respectively. Notice also that for all gauge groups the minimum is achieved for charged fermions only. The presence of charged scalars systematically enhances $C>C_{\rm min}$.
Thus, coming back to the scalar self interactions, even in the most favourable scenario where the one-loop coefficient vanishes and  the gauge coupling is perturbatively small, a cancellation between the two~loop gauge and the three or four~loop scalar contributions requires scalar couplings of order unity owing to the lower bounds \eq{Cbound}, \eq{Cmin}.\footnote{ For this estimate we have assumed that the relevant loop factor $(4\pi)^2$ is scaled into the definition of the scalar self-coupling, consistent with our conventions for the gauge and Yukawa couplings.}   
Hence, the feasibility of such a scenario  necessitates non-perturbatively large scalar couplings, outside the perturbative domain. 
We conclude that non-abelian gauge theories with any type of self interacting scalar matter, and with or without fermionic matter but without Yukawa interactions,  cannot become asymptotically safe within perturbation theory. This result also completes the no go theorems stated in Tab.~\ref{tFinal}~{\bf b)} and~{\bf c)}  in the presence of scalar matter.\step

{\bf 7.} 
Yukawa couplings are naturally present in settings with both scalar and fermionic matter fields \cite{Yukawa:1935xg}, and contribute to the running of (some of) the gauge couplings provided that (some of) the fermions carry charges under (some of) the gauge groups. Scalars may or may not carry charges.
Yukawa couplings are technically natural \cite{'tHooft:1979bh} and cannot be switched-on by fluctuations: the limit of  vanishing Yukawa couplings constitutes an exact fixed point of the theory. 
\step

For concreteness we consider simple non-abelian or abelian gauge theories with the most general Yukawa interactions taking the form $\sim\frac12({\bf Y}^A)_{JL} \phi^A \,\psi_J \,\zeta\,\psi_L$ with $\zeta=\pm i \sigma_2$, with Weyl indices suppressed. In perturbation theory the Yukawa couplings ${\bf Y}^A$ contribute to  the renormalisation of the gauge coupling starting at the two loop level, and the beta function \eq{eq:GaugeBeta}  is replaced by \cite{Machacek:1983tz}
\bea\label{Yukawa}
\beta&=& \alpha^2\left(-B\, +C\,\alpha 
- 2\,Y_4
\right) \,.
\eea
The Yukawa couplings enter  through the new term $Y_4=\Tr [{\bf C}^F_2\, {\bf Y}^A\,({\bf Y}^A)^\dagger]/d(G)$, with $d(G)$  the dimension of the gauge group, ${\bf Y}^A$ the (matrix of) Yukawa couplings, ${\bf C}^F_2$ the matrix of quadratic Casimirs of the fermionic irreps, and the trace summing over all fermionic indices. 
Notice that we have  scaled the loop factor of $(4 \pi)$ into the definition of ${\bf Y}^A$.
The coefficients $B$ and $C$ are as in \eq{eq:OneLoop} and \eq{eq:TwoLoop}. 
In general, the matrix ${\bf C}^F_2$ is diagonal according to the fermionic irreps, implying that $Y_4$ is positive as long as (some of) the Yukawa couplings are non-vanishing. Positivity of $Y_4$ can be made manifest by rewriting it as
\beq
\label{pos}
Y_4=\sum_{AJL}\,{S^{F_J}_2}\left|({\bf Y}^A)_{JL}\right|^2/{d(F_J)}\ge 0\,.
\eeq
It follows that Yukawa couplings  contribute with an overall negative sign to the running of gauge couplings, irrespective of the sign of the one~loop gauge coefficient $B$. 
Assuming that the Yukawa couplings, and thus $Y_4$, take a fixed point of their own,  interacting fixed points of \eq{Yukawa} take  the form \eq{BC} except that the one loop coefficient is effectively  shifted $B\to B'=B+2\,Y^*_4$, with
\beq B'\ge B\,.
\eeq
This Yukawa-induced shift has important implications.  Most notably,
in settings where the gauge sector is asymptotically non free ($B\le 0$), the Yukawa contribution $Y_4^*$ may effectively change the sign of the one loop 
coefficient  $(B'>0)$, thereby enabling a viable interacting  fixed point 
\beq \label{B'C}
\alpha_*=\frac{B'}{C}\,.
\eeq
In more physical terms, for infrared free theories these findings state that the growth of the gauge coupling with energy, as dictated by the positive one and two~loop gauge contributions \eq{nogo}, is invariably slowed down, and, as long as $B'>0$, eventually brought to a halt by  Yukawa interactions. In particular, the occurrence of a UV Landau pole in the gauge coupling can be avoided dynamically.
As we have  shown earlier, neither scalar self interactions nor further gauge couplings are able to negotiate a fixed point at weak coupling once $B\le 0$. We therefore conclude that Yukawa interactions are the {\it only} type of interactions that can generate an interacting UV fixed point for {\it any} weakly coupled gauge theory.
\step

In view of the above it is useful to investigate the Yukawa sector in more detail.
To that end, we exploit
the explicit flow for the Yukawa couplings $\bm\beta^A=d {\bf Y}^A/d\ln \mu$. At the leading non-trivial order in perturbation theory which is one loop,
it takes the form \cite{Coleman:1973sx,Cheng:1973nv}
\bea\label{BetaYukawa}
	{\bm\beta}^A&=&{\bf E}^A(Y)-
 \alpha\,{\bf F}^A(Y)\,.
 \eea
The terms ${\bf E}^A(Y)$, which  are of cubic order in the Yukawa couplings,  arise from fluctuations of the fermion and scalar fields and encode vertex and propagator corrections \cite{Coleman:1973sx}.  General expressions for ${\bf E}^A$ in the conventions adopted here are given in  \cite{Machacek:1983fi,Luo:2002ti}.  The terms  ${\bf F}^A(Y)=3  \{{\bf C}_2^{F},{\bm Y^A}\}$ originate primarily from gauge field fluctuations and are  (block-)diagonally proportional to $\bm Y^A$ following the fermion irreps \cite{Cheng:1973nv}.  Scalar self couplings contribute to \eq{BetaYukawa} starting at two loop and can be neglected for sufficiently small couplings. 
\step

The nullcline condition $\bm\beta^A(Y,\alpha)=0$ for the Yukawa couplings  has two types of solutions. The Gaussian fixed point ${\bf Y}_*^A=0$ always exists, because both ${\bf E}^A$ and ${\bf F}^A$ vanish individually for vanishing Yukawa couplings, whence  $\bm\beta^A(Y=0,\alpha)=0$. In addition, and provided that the gauge coupling is non-vanishing, the two terms in \eq{BetaYukawa} can balance against each other. Dimensional analysis  shows that the functions $\bar{\bm\beta}^A(C)\equiv {\bm\beta}^A(\sqrt{\alpha}\, C,\alpha)/\alpha^{3/2}$ are independent of the gauge coupling $\alpha$, implying that Yukawa nullclines take the form 	
\beq\label{YukawaFP}
{\bf Y}^A_*=\frac{g}{4\pi}\, {\bf C}^A\,.
\eeq
The ``reduced'' Yukawa couplings ${\bf C}^A$ are numerical matrices independent of the gauge coupling $g$ which solve $\bar{\bm\beta}^A(C)=0$, meaning ${\bm E}^A(C)={\bf F}^A(C)$ for ${\bf C}^A\neq 0$. 
Evidently ${\bf C}^A=0$ corresponds to the Gaussian.\footnote{For any nullcline ${\bf C}^A$ \eq{YukawaFP}, $-{\bf C}^A$ and ${\bf C}^{A\dagger}={\bf C}^{A\,*}$ are physically equivalent nullclines. In the literature one-loop
nullclines are sometimes referred to as ''fixed points'' (for the reduced couplings) or ``eigenvalue conditions'' \cite{Chang:1974bv}.} The solutions \eq{YukawaFP} are promoted to genuine fixed points of the coupled system \eq{Yukawa}, \eq{BetaYukawa}  iff the gauge coupling simultaneously takes a real  fixed point $g_*$ \eq{B'C}. At the fixed point, perturbativity in the Yukawa couplings then follows parametrically from 
perturbativity 
in the gauge coupling.
\step

Inserting the nullcline back into \eq{Yukawa} we find that the Yukawa-induced terms are of order $\alpha^3$ owing to  
\eq{YukawaFP}. This  establishes that the shifted one loop coefficient $B'$ depends linearly on $\alpha$ through $Y_4^*$, meaning that \eq{B'C} constitutes an implicit equation for $\alpha_*$. The implicit dependences are resolved by accounting for the Yukawa contributions as, effectively, modifications of the two loop coefficient. We find 
\beq\label{D}
Y_4=\dy\cdot\alpha
\eeq
where the coefficient $\dy=\Tr [{\bf C}^F_2\, {\bf C}^A\,({\bf C}^A)^\dagger]/d(G) \ge 0$ only depends on group theoretical weights and the reduced Yukawa couplings parametrising the nullcline, but not on the gauge coupling.  The projection of the flow for the gauge coupling \eq{Yukawa} along a hypersurface with $\bm\beta^A=0$ then takes  the form \eq{eq:GaugeBeta} except that the two loop gauge coefficient $C$ is shifted into $C\to C'=C- 2\,\dy$. The shift term vanishes iff all Yukawa couplings vanish but is strictly negative otherwise, whence
\beq \label{lower} 
C'\le C\,.
\eeq
This result makes it manifest that Yukawa contributions  can dynamically lower the effective two loop coefficient, possibly avoiding the no go theorem \eq{nogo}. Furthermore, the shift \eq{lower} implies that interacting fixed points for the gauge coupling take  the form \eq{BC} with $C\to C'$,
\beq\label{BC'}
\alpha_*=\frac{B}{C'}\,.
\eeq
We stress that the expressions \eq{B'C} and \eq{BC'} for the gauge coupling fixed point are equivalent and numerically identical. For practical purposes, however, the latter representation, if available, is preferred as it provides the fully resolved version of the former. 
Following on from our earlier discussion, the fixed points \eq{BC'} are physical as long as $B\cdot C'>0$, and perturbative if $|B|\ll |C'|$. If $B>0$ and $C'>0$, they constitute infrared fixed points of the theory, similar to Banks-Zaks fixed points except for the additional presence of Yukawa interactions. If $B<0$ and $C'<0$, they constitute interacting UV fixed points and qualify as asymptotically safe UV completions for the theory, see Tab.~\ref{tFP}~{\bf c)} for a summary. No such weakly coupled UV completion can arise without Yukawa interactions.\step

\begin{table}[t]
\begin{center}
\begin{tabular}{GccccG}
\colb
\toprule
&&&&&\\[-3.5mm]
 \colb
&&&& \multirow{ 1}{*}{\ \bf asymptotic\ }&
\\ 
\colb
\multirow{ -2}{*}{\ case \ }&
\multirow{ -2}{*}{gauge group} &
\multirow{ -2}{*}{matter}&
\multirow{ -2}{*}{\ \bf Yukawa \ }& 
\multirow{ 1}{*}{\ \bf safety\ }&
\multirow{ -2}{*}{info} 
\\[.4mm]
\midrule
\rowcolor{LightGray}
&&&&\coln&\\[-3.5mm]
\rowcolor{LightGray}
&&&&\coln&\\[-3mm]
\rowcolor{LightGray}
{\bf a)}& simple&fermions in irreps&\bf  No &\coln \bf No & Ref.~\cite{Caswell:1974gg}
\\[.4mm]
\midrule
\cellcolor{white}
&&&&\coln&\cellcolor{white}
\\[-3mm]
\cellcolor{white}
&&fermions, any rep&\bf  No& \coln \bf No &\cellcolor{white}
 $\eq{nogo}$
\\
\cellcolor{white}
{\bf b)}&  simple or abelian  &scalars, any rep&\bf  No& \coln \bf No & \cellcolor{white}
$\eq{nogo}, \eq{Cmin}$
\\
\cellcolor{white}
&  
& \ fermions and scalars, any rep \ &\bf  No& \coln \bf No &\cellcolor{white}
 $\eq{nogo}, \eq{Cmin}$
\\[1mm]
\midrule
\rowcolor{LightGray}
&&&&\coln&\\[-3mm]
\rowcolor{LightGray}
& semi-simple,&  fermions, any rep &\bf  No &\coln \bf No & $\eq{nogoa}$
\\[.4mm]
\rowcolor{LightGray}
{\bf c)}&with or without  &   scalars, any rep &\bf  No &\coln \bf No & $\eq{nogoa}, \eq{Cmin}$
\\
\cellcolor{LightGray}
& \cellcolor{LightGray}
 abelian factors&
\cellcolor{LightGray}fermions and scalars, any rep &\cellcolor{LightGray}\bf  No &\coln \bf No & $\eq{nogoa}, \eq{Cmin}$
\\[1mm]
\midrule
\cellcolor{white}
&&&&\coly&\cellcolor{white}
\\[-2.5mm]
\multirow{1}{*}{\cellcolor{white}
{\bf d)}}&simple or abelian &
\multirow{1}{*}{fermions and scalars, any rep}&
\multirow{ 1}{*}{\bf Yes} &
\multirow{1}{*}{\coly \bf Yes}${}$  & 
\cellcolor{white}$\eq{BC'}, 
\eq{scalarC}$
\\[.5mm]
\midrule
\rowcolor{LightGray}
&&&&\coly&
\\[-3.5mm]
\rowcolor{LightGray}
&semi-simple, with or  &&&\coly &\\ 
\rowcolor{LightGray}
\multirow{-2}{*}{{\bf e)}}& 
without abelian factors& 
\multirow{ -2}{*}{fermions and scalars, any rep}&
\multirow{ -2}{*}{\bf Yes}   &  
\multirow{ -2}{*}{\coly \bf Yes}   &
\multirow{ -2}{*}{\cellcolor{LightGray}$\eq{FPa'b}, \eq{scalarC}$}
	\\  
	\bottomrule
\end{tabular}
\end{center}
 \caption{Asymptotic safety in  gauge theories coupled to matter with
{\bf a)} -- {\bf c)} stating strict no go theorems and {\bf d)} -- {\bf e)} necessary and sufficient conditions.
 \label{tFinal}}
\end{table}

We conclude that Yukawa couplings offer a dynamical mechanism to negotiate  interacting fixed points in gauge theories.  Most importantly, for asymptotically non-free gauge theories with $B\le 0$, they offer a {\it unique} mechanism to generate weakly interacting fixed points.
The strict no~go~theorem \eq{nogo} may then be circumnavigated under the auxiliary condition that the Yukawa-induced shift term comes out large enough for $C'$ to turn negative. This result,  summarised in Tab.~\ref{tFinal}`{\bf d)}, thus takes the form of a necessary condition for asymptotic safety. \step

{\bf 8.} Our results are straightforwardly generalised to gauge-Yukawa theories with several abelian or non-abelian gauge group factors, assuming that some or all of the fermions are charged under some or all of the gauge groups, while the scalars may or may not be charged. The renormalisation of the gauge couplings then takes the form \cite{Machacek:1983tz}
\bea\label{gaugeYukawa}
\beta_a&=& \alpha_a^2\left(-B_a\, +C_{ab}\,\alpha_b 
- 2\,Y_{4,a}
\right) \,,
\eea
where the two loop Yukawa contributions now arise through
$Y_{4,a}=\Tr [{\bf C}^{F_a}_2\, {\bf Y}^A\,({\bf Y}^A)^\dagger]/d(G_a)\ge 0$. 
As is evident from the explicit expression, the quadratic Casimir of the fermions takes the role of a projector to identify  the contributions to the running of $\alpha_a$. The running of the Yukawa couplings continues to be given by \eq{BetaYukawa}, except that further  gauge field contributions turn the last term into a sum over gauge groups $\alpha\, {\bf F}^A\to \alpha_a\, {\bf F}_{a}^A$ with ${\bf F}_a^A(Y)=3  \{{\bf C}_2^{F_a},{\bm Y^A}\}$~\cite{Machacek:1983fi}. This modification leads to a larger variety of Yukawa nullclines, depending on which of the gauge couplings take vanishing or non-vanishing values at the fixed point.  Provided that some or all of the Yukawa couplings take interacting fixed points they will contribute to the running of the gauge couplings \eq{gaugeYukawa}  through $Y_{4,a}^*\ge 0$. Consequently, the gauge beta functions reduce to  the form \eq{eq:MultiGaugeBeta} except that the one loop coefficients are effectively shifted,
$B_a\to B'_a=B_a+ 2 \,Y_{4,a}^*$, due to the fixed point in the Yukawa sector. Most importantly, we observe that
\beq\label{B'a}
B'_a \ge B_a\,.
\eeq
Equality holds true iff all Yukawa couplings take Gaussian values.  The shift \eq{B'a} implies that gauge coupling fixed points of the theory  arise as the solutions of
\beq\label{FPa'b}
B'_a=C_{ab}\,\alpha_b^* \,,
\quad
{\rm subject\ to}\quad \alpha^*_b\ge 0\,.
\eeq
Once more, this structure has important implications. Following on from our earlier discussion of \eq{FPab}, the fixed point condition \eq{FPa'b} can have  physical solutions iff all $B'_a$ are positive. Due to \eq{B'a} this is naturally the case as long as  each gauge group factor is asymptotically free. The theory is then asymptotically free in all gauge factors with interacting fixed points of the Banks-Zaks and the gauge-Yukawa type, and combinations and products thereof. The decisive  difference with \eq{FPab} comes into  its own for theories where some or all $B_a$ are negative. 
Provided that the Yukawa-induced shift terms ensure that all $B'_a$ become positive numbers even if one or several of the gauge factors are not asymptotically free, the fixed point condition \eq{FPa'b} can have a variety of novel  solutions,
see Tab.~\ref{tFP}~{\bf d)}. Such fixed points are genuinely of the gauge-Yukawa type, and furthermore constitute candidates for  asymptotically safe UV completions of the theory. 
Also, no such fixed point can  arise out of theories with \eq{FPab}, which once more highlights the pivotal role played by  Yukawa interactions. \step

As a final remark, we note that the fixed point condition \eq{FPa'b} still depends implicitly on the gauge couplings through $B'_a$, once $Y_4$ is evaluated on a nullcline. 
It is straightforward to resolve the implicit dependence provided that  $Y_{4,a}$  takes the form
\beq\label{Dab}
Y_{4,a}=\dy_{ab}\,\alpha_b
\eeq 
along Yukawa nullclines, in analogy to \eq{D}.\footnote{The form \eq{Dab} is evident 
if only one of the gauge couplings, say $g_b$, is non-vanishing. The  
nullcline takes the  form 
${\bf Y}^A_{b,*}=\frac{g_b}{4\pi}\, {\bf C}_b^A$, see \eq{YukawaFP}, 
with $\bm C^A_b$  a solution of ${\bf E}^A(C)={\bf F}_b^A(C)$, leading to 
$\dy_{ab}=\Tr [{\bf C}^{F_a}_2\, {\bf C}_b^A\,({\bf C}_b^A)^\dagger]/d(G_a)\ge 0$.
More generally,  \eq{Dab} holds true for any quantum field theory whose  one loop Yukawa vertex corrections obey ${\bf Y}^B{\bf Y}^{\dagger A}{\bf Y}^B={\bf Y}^{A}\,\Tr\, {\bf M}^{BC}({\bf Y}^{\dagger B}{\bf Y}^C+{\bf Y}^{\dagger C}{\bf Y}^B)$ for some matrix $({\bf M}^{BC})_{JL}=m^B_J\,\delta^{BC}\,\delta_{JL}$ which is block-diagonally proportional to the identity in  field space with real  $m^B_J$.
In these cases 
the flow for the Yukawa couplings \eq{BetaYukawa} are mapped explicitly onto 
closed flows for their squares $|({\bf Y}^{A})_{JK}|^2$  whose nullclines, and consequently $Y_{4,a}$ on nullclines, are linear functions of the squares of the gauge couplings, $\alpha_b$.
In  theories with more complex Yukawa vertex corrections  
($e.g.$~Pati-Salam, trinification)
the relation between $Y_{a,4}$ and $\alpha_b$ takes a more general form.} Continuity in each of the gauge couplings $\alpha_b\ge 0$ together with the non-negativity of $Y_{4,a}$ allows us to observe that the matrix  $(\dy_{ab})$ is  non-negative.  
The flow of the gauge couplings  \eq{gaugeYukawa} is reduced to \eq{eq:MultiGaugeBeta}, except that the two loop term is shifted $C_{ab}\to C'_{ab}=C_{ab}-2\,\dy_{ab}$ following \eq{Dab}.  We conclude that the Yukawa contributions along nullclines effectively reduce the two loop gauge contributions to the renormalisation of  gauge couplings. 
In this representation, the fixed point condition \eq{FPa'b} turns into the equivalent form
\beq\label{FPab'}
B_a=C'_{ab}\,\alpha_b^* \,,
\quad
{\rm subject\ to}\quad \alpha^*_b\ge 0\,.
\eeq
For non-negative $C'_{ab}$, as has been shown above, interacting fixed points can only be realised if all gauge group factors are asymptotically free. Here, however,  the matrix $(C'_{ab})$ is no longer required to be strictly non-negative, unlike the matrix $(C_{ab})$ of two loop gauge contributions, and the no go theorem \eq{nogoa} can be avoided owing to the Yukawa contributions.
In view of the asymptotic safety conjecture, this completes our proof that charged fermions with charged or uncharged scalars and, most crucially, Yukawa interactions, constitute  strictly  necessary ingrediences  for interacting UV fixed points  in general weakly coupled gauge theories, see Tab.~\ref{tFinal}~{\bf e)}.\step

\begin{table}[t]
\begin{center}
\begin{tabular}{ccccccc}
  \rowcolor{LightGreen}
\toprule
&&&&&&\\
 \rowcolor{LightGreen}
\multirow{-2}{*}{case}&\multirow{-2}{*}{gauge group}&\multirow{-2}{*}{Yukawa\ }&\multirow{-2}{*}{  \ \ parameter\ \ \ }&\multirow{-2}{*}{ \  interacting FPs \   }&\multirow{-2}{*}{type} 
& \multirow{-2}{*}{info}
\\[.4mm]
\rowcolor{LightGray}
\midrule
&&&&\cchi&&\\[-3mm]
\rowcolor{LightGray}
\multirow{1}{*}{{\bf a)}}& \multirow{1}{*}{simple}
& \bf No &$B>0$ and $C>0$\ \ \ \ \ \quad\quad& \cchi Banks-Zaks  
&\bf  IR & Refs.~\cite{Caswell:1974gg,Banks:1981nn}
\\[.45mm]
\midrule
\rowcolor{white}
&&&&\cchi&&
\\[-3mm]
& semi-simple,&&&\cchi Banks-Zaks and& &
\\
\multirow{-2}{*}{{\bf b)}}& no $U(1)$ factors &\multirow{-2}{*}{\bf No} & \multirow{-2}{*}{all $B_a>0$}
& \multirow{-1}{*}{\cchi products thereof}
&\multirow{-2}{*}{\bf IR} &\multirow{-2}{*}{soln of \eq{FPab}}
 \\[.4mm]
\midrule
\\[-4.1mm]
\rowcolor{LightGray}
& &&&\cchi&&\\[-3mm]
\rowcolor{LightGray}
&\multirow{1}{*}{simple}
&\bf  Yes &$B>0$ and  $C>0>C'$\ \ &\cchi Banks-Zaks &\bf  IR& Fig.~\ref{pAFb}
\\[.8mm]
\cellcolor{LightGray}
&\cellcolor{LightGray}\cellcolor{white}
&\cellcolor{white}
&&\cchi&&\\[-3mm]
\cellcolor{LightGray}
{\bf c)}
&\cellcolor{LightGray}\multirow{1}{*}{\cellcolor{white}simple}
&\bf  Yes &$B>0$ and $C>C'>0$\ \ &\cchi BZ 
and GYs 
& \bf IR
&
Fig.~\ref{pAFc}
\\[.8mm]
\rowcolor{LightGray}&\cellcolor{LightGray}&&&\cchi&&\cellcolor{LightGray}\\[-3mm]
\rowcolor{LightGray}& \ simple or abelian \ &\bf  Yes&$B<0$ and $C'<0$ \ \ \ \quad\quad& 
\cchi gauge-Yukawas & \bf UV/IR  & Fig.~\ref{pAFd}
\\[.5mm]
\midrule
&&&&\cchi&&
\\[-3mm]
& semi-simple, with &&&
\cchi BZs and GYs and&&\multirow{-2}{*}{}
\\
\multirow{-2}{*}{{\bf d)}}& or without $U(1)$ factors &\multirow{-2}{*}{\bf Yes}  &\multirow{-2}{*}{all $B'_a>0$}    & 
\multirow{-1}{*}{\cchi products thereof} 
&\multirow{-2}{*}{\bf UV/IR}&\multirow{-2}{*}{
soln of \eq{FPa'b}}
\\[.4mm]
\bottomrule
\end{tabular}
\end{center}
 \caption{Summary of weakly interacting fixed points in gauge theories, detailing the availability of Banks-Zaks (BZ) or gauge-Yukawa (GY) type fixed points, or combinations and products thereof.
 \label{tFP}}
\end{table}

{\bf 9.} Gauge-Yukawa fixed points necessitate  scalar fields. 
Consequently, two auxiliary conditions arise: Firstly, the scalar sector must achieve a   fixed point of its own, interacting or otherwise. Secondly, the scalar sector must admit a stable ground state. 
To appreciate that both of these requirements are non-empty, we consider the renormalisation group flow  $\bm{\beta}=d {\bm \lambda}/d\ln \mu$ for the quartic scalar couplings $\bm \lambda = (\lambda_{ ABCD})$ based on the interaction Lagrangean $\sim \frac1{4!}\lambda_{ ABCD}\, \phi^A\phi^B\phi^C\phi^D$. To leading order the beta functions $\bm \beta=\bm \beta(\lambda,Y,\alpha)$ depend quadratically  on the quartics, on the Yukawa and gauge couplings, and on group theoretical factors related to the gauge transformations of the scalars (if charged) \cite{Cheng:1973nv}. Explicit expressions and generalisations for product gauge groups can be found in \cite{Machacek:1984zw,Luo:2002ti}.  
Scalar self couplings are not technically natural \cite{'tHooft:1979bh} and
can be switched-on by fluctuations of the fermions (due to the presence of Yukawa couplings) or by fluctuations of the gauge fields (if the scalars are charged), 
implying that $\bm \beta(\lambda=0,Y,\alpha)\neq 0$ in general.  
\step

Next we turn to the scalar nullclines $\bm \beta=0$, subject to $\bm \beta^A\to 0$. 
Using dimensional analysis, we observe that the functions $\bar{\bm\beta}(\bar {C}, C)\equiv \bm \beta(\alpha\, \bar{C},\alpha\, C,\alpha)/\alpha^2$ are $\alpha$-independent. 
 The  implicit solutions $\bar{\bm C}$ of  the quadratic algebraic equations $\bar{\bm\beta}(\bar C(C),C)=0$ provide us with 
 \beq\label{scalar}
 \bm \lambda_*=\alpha \, \bar{\bm C}\,.
  \eeq
The ``reduced'' scalar couplings $\bar{\bm C}$ are numerical tensors which depend on group theoretical factors and the reduced Yukawa couplings, but not explicitly on the gauge coupling. 
Since the quartics do not impact on the gauge-Yukawa flow (to leading order) it is immaterial for this analysis whether the gauge coupling is slowly running or sitting on a fixed point.
\step

Qualitatively and quantitatively  different types of solutions $\bm \lambda_*$ arise for all physically inequivalent  Yukawa nullclines with $\bm C^A\neq0$, and with $\bm C^A\to 0$.
In either of these cases, owing to the quadratic nature of the defining equations,  solutions 
\eq{scalar}
generically come up in inequivalent pairs $\bar{\bm C}_\pm$ per Yukawa nullcline with 
complex entries. 
Reality of quartic  couplings is not automatically guaranteed and must be required as an auxiliary condition.
Vacuum stability necessitates that $\bm \lambda_*$ is a positive-definite tensor.\footnote{In the presence of flat directions,  Coleman-Weinberg type resummations \cite{Coleman:1973jx} for the leading logarithmic corrections of the effective potential will have to be invoked \cite{,Gross:1973ju}.} 
This information is not encoded in the renormalisation group flow even if the scalar couplings come out real, meaning that the stability of the  effective potential $V_{\rm eff}(\phi)$ provides an independent constraint. We therefore conclude that  \eq{scalar}, subject to
\beq\label{scalarC}
\lambda^*_{ ABCD}={\rm real}\,,\quad{\rm and}\quad V_{\rm eff}(\phi)={\rm stable}\,,
\eeq
are  mandatory auxiliary conditions for gauge theories with scalar matter to display a physically acceptable scalar sector, in addition to the conditions for free or interacting fixed points in the gauge or gauge-Yukawa 
sectors. 

A few comments are in order: $(i)$ Solutions of \eq{scalarC} with $\bm C^A\neq 0$ are mandatory for gauge-Yukawa fixed points and  for asymptotic safety \cite{Litim:2014uca,Litim:2015iea}. Those with $\bm C^A= 0$ are mandatory for Banks-Zaks fixed points in the presence of scalar matter.
$(ii)$   Both of \eq{scalarC} must be imposed irrespective of the UV or IR nature of the underlying fixed point. 
$(iii)$ If two solutions $\bar{\bm C}_\pm$ are physical, one of them is UV and the other IR relevant. 
$(iv)$ Solutions to \eq{scalar}, \eq{scalarC} also control trajectories in the vicinity of free or interacting fixed points
\cite{Chang:1974bv}.
Those with $\bm C^A\neq 0$ entail that gauge, Yukawa, and scalar couplings run at the same rate and govern the approach to gauge-Yukawa fixed points. Those with $\bm C^A\to 0$ (referring to reduced Yukawa couplings  which approach the Gaussian very rapidly $\bm Y^A(\alpha)/ \sqrt \alpha\equiv \bm C^A(\alpha)\ll 1$)  are relevant for asymptotically free theories  to display complete asymptotic freedom, and for trajectories approaching Banks-Zaks fixed points.
Scalar couplings  then run  into the Gaussian UV fixed point either alongside the gauge coupling, or faster $\bm \lambda_*(\alpha)/\alpha\ll 1$. The latter follows from  the  $\alpha$-dependence of the reduced Yukawa couplings $\bm C^A(\alpha)$ which entails an implicit $\alpha$-dependence for the quartics \cite{Cheng:1973nv}.
$(iv)$  A  method to find solutions 
in the limit $\bm C^A\to 0$ has been detailed in \cite{Bais:1978fv}. Physical solutions for the combined Yukawa and scalar nullclines  with \eq{scalarC} exist and are  known 
 for  a number of theories \cite{Ma:1974cm,Fradkin:1975rz,Kalashnikov:1976hr,Callaway:1988ya}.\footnote{See  \cite{Litim:2014uca,Litim:2015iea} and  \cite{Giudice:2014tma,Holdom:2014hla} for recent results in the context of asymptotic safety and  asymptotic freedom, respectively.} 
 \step

  This completes the derivation of necessary and sufficient conditions of existence for weakly interacting fixed points in general gauge theories coupled to matter. \step

\begin{figure}[t]
\begin{center}
\includegraphics[scale=.66]{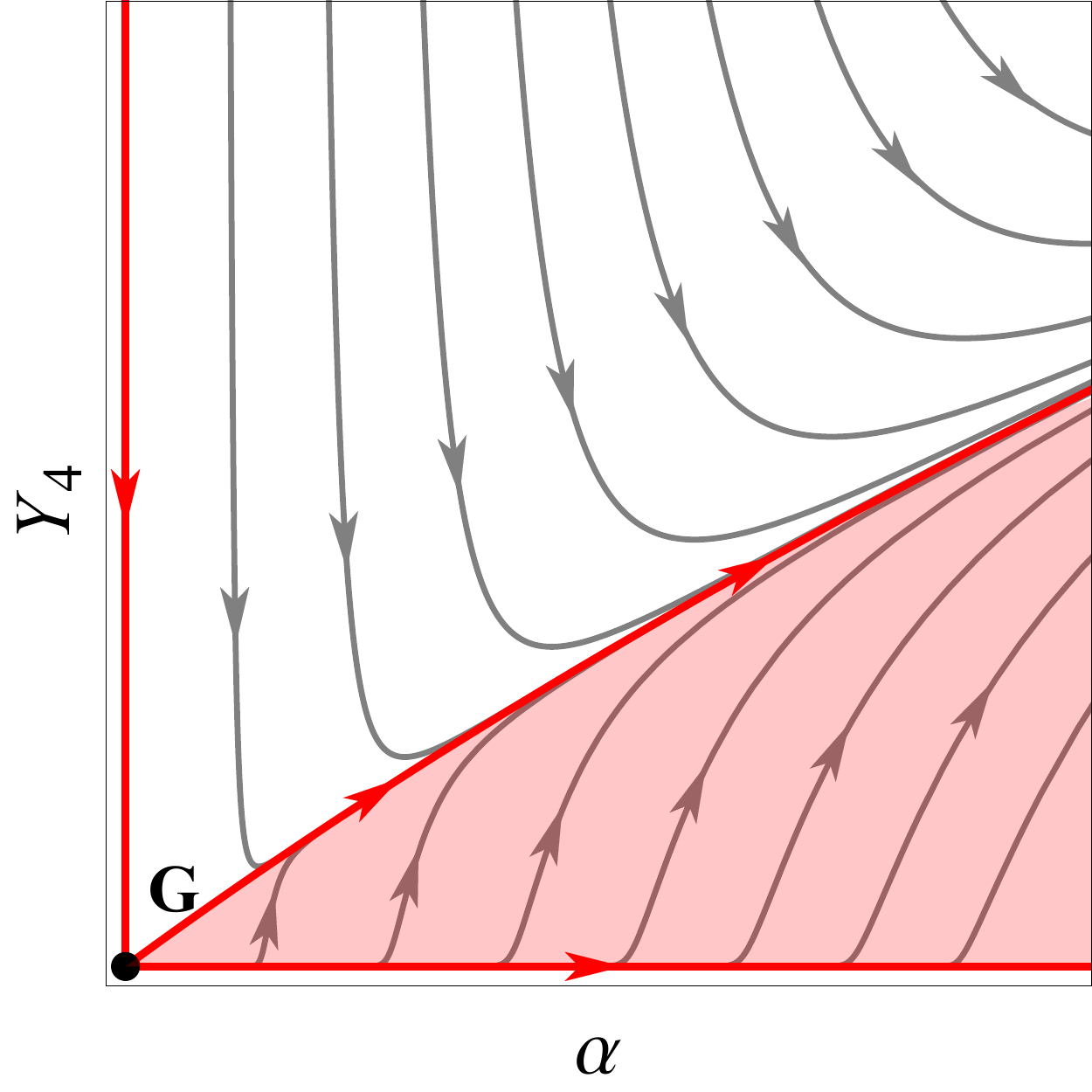}
\caption{Phase diagram of  gauge-Yukawa theories with $B>0$ and $C<0$ at weak coupling 
showing asymptotic freedom and the Gaussian UV fixed point (G). Arrows indicate the flow towards the IR. The red-shaded area covers the set of UV complete trajectories emanating form the Gaussian UV fixed point. The Yukawa nullcline acts on trajectories as an IR attractor.
}\label{pAFa} 
\end{center}
\end{figure}

\begin{figure}[t]
\begin{center}
\includegraphics[scale=.66]{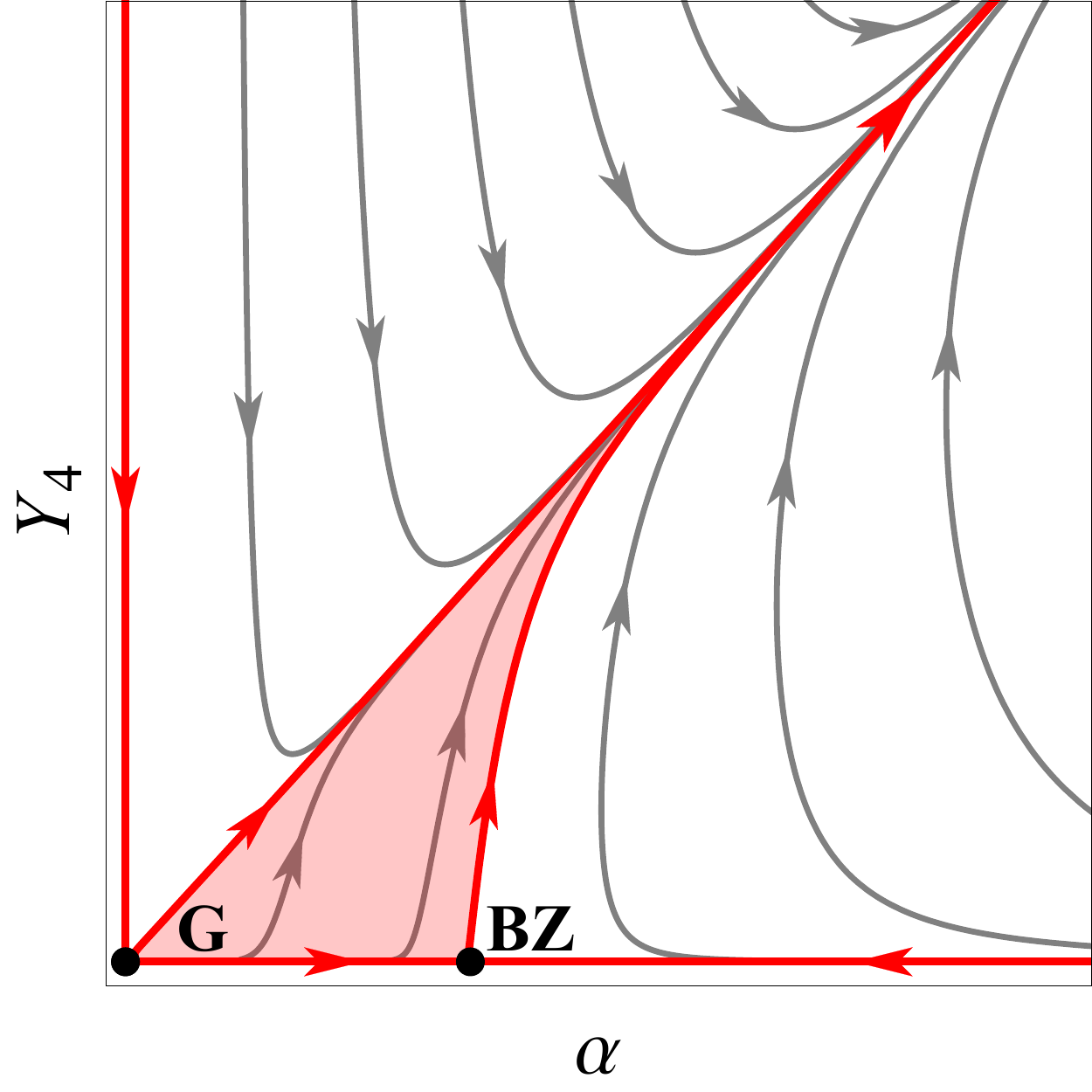}
\caption{Phase diagram of  gauge-Yukawa theories with $B>0$ and $C>0>C'$ at weak coupling showing asymptotic freedom with the Gaussian and the Banks-Zaks fixed point (BZ). 
Notice the funneling of all UV free trajectories towards the Yukawa nullcline as furthered by the Banks-Zaks fixed point. 
}\label{pAFb} 
\end{center}
\end{figure}

\begin{figure}[t]
\begin{center}\includegraphics[scale=.66]{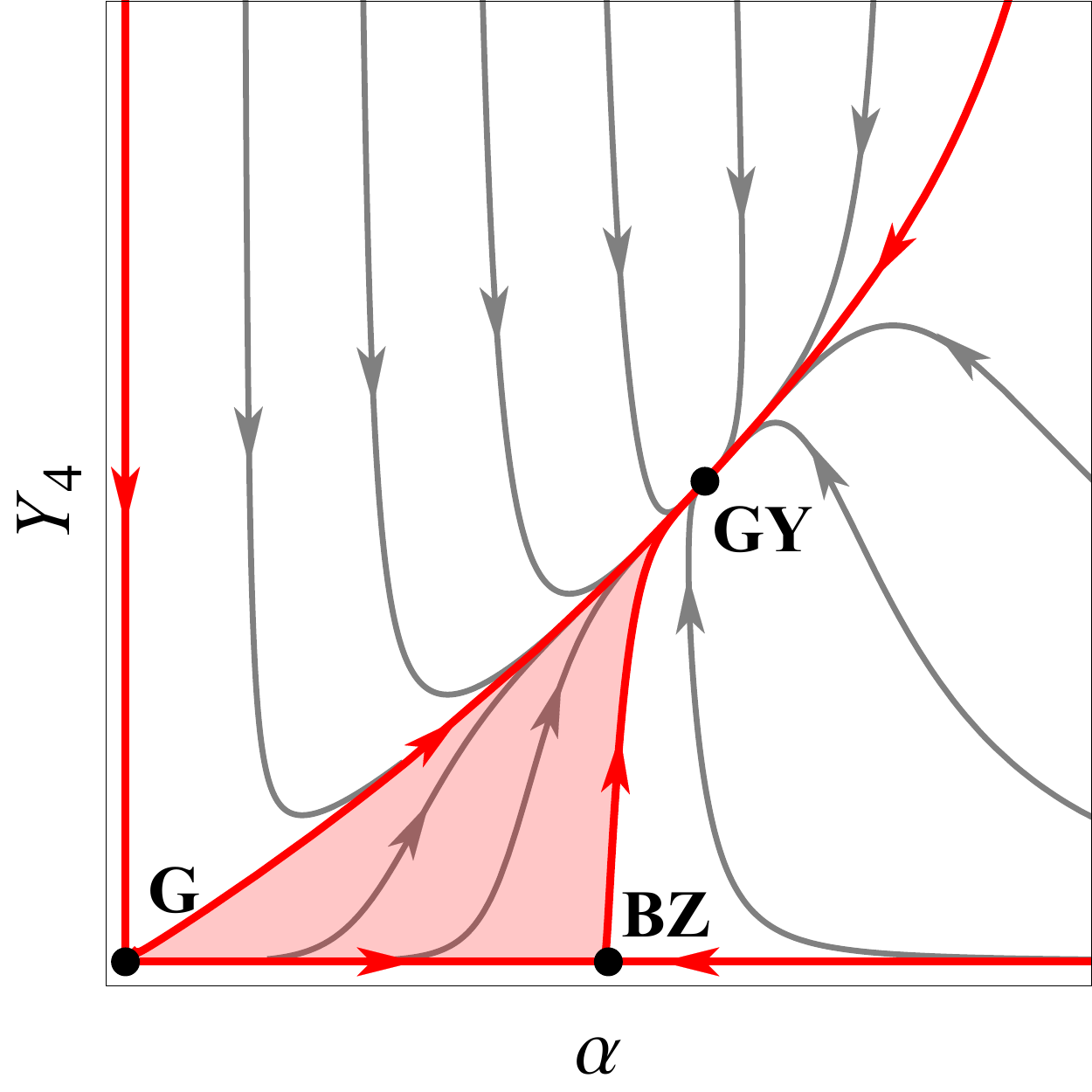}
\caption{Fixed points and phase diagrams of gauge-Yukawa theories with $B>0$ and $C>C'>0$ at weak coupling 
showing asymptotic freedom with Gaussian, Banks-Zaks, and gauge-Yukawa  fixed points (GY).
Notice that the gauge-Yukawa fixed point attracts UV free trajectories emanating from the Gaussian.}\label{pAFc} 
\end{center}
\end{figure}

\begin{figure}[t]
\begin{center}
\includegraphics[scale=.66]{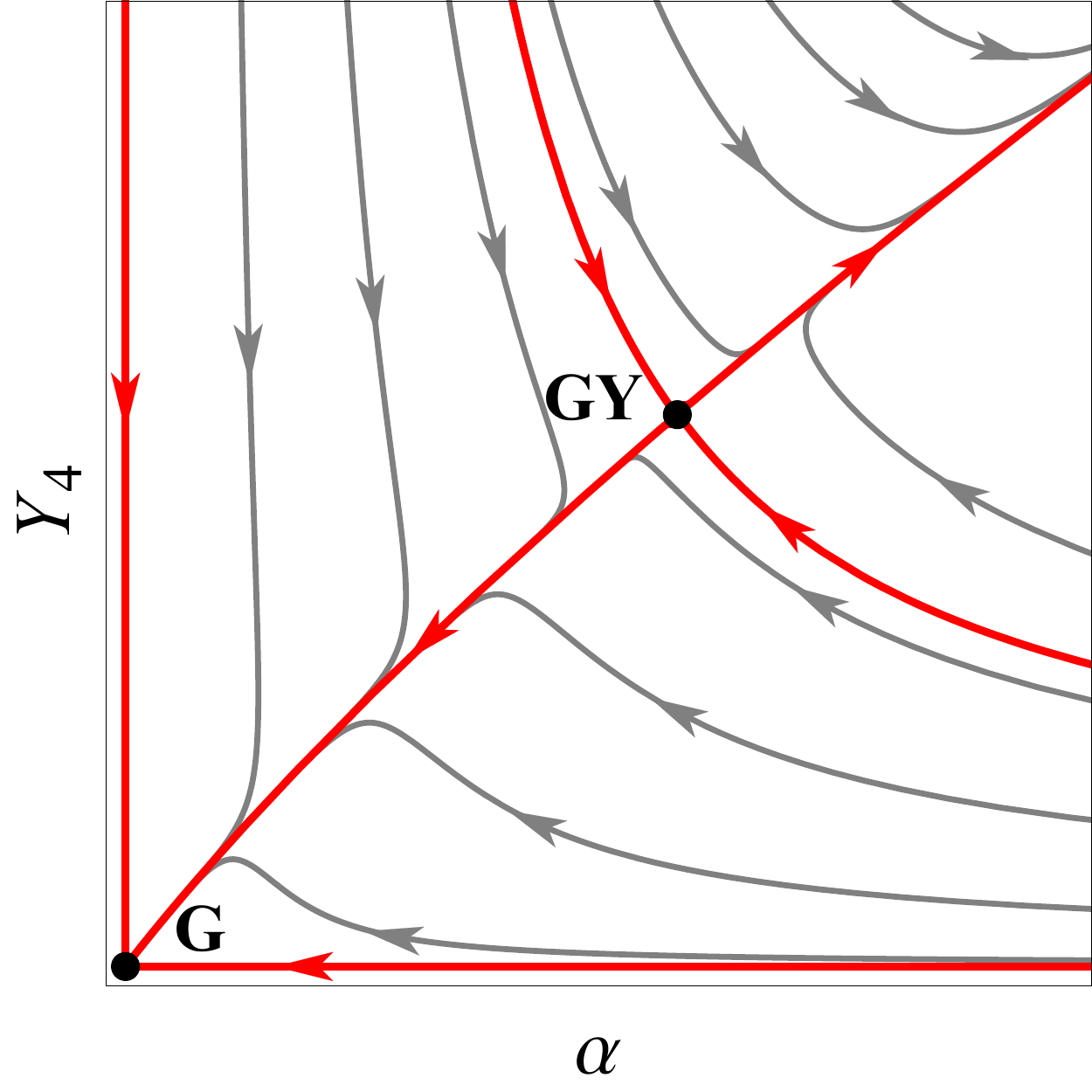}
\caption{Fixed points and phase diagrams of gauge-Yukawa theories with $B<0$ and $C'<0$ at weak coupling showing  asymptotic safety together with the Gaussian and gauge-Yukawa fixed points. Notice that the set of UV finite trajectories is confined to a hypercritical surface dictated by the Yukawa nullcline.}\label{pAFd} 
\end{center}
\end{figure}

{\bf 10.}   Next, we  return to the starting point of our investigation where we observed that the competition between gauge field and matter fluctuations, and hence the relative signs and size of the loop coefficient $B$ and $C$ (for theories with a simple gauge group) determines the fixed point structure. However, it has become clear that a third quantity, $C'$, controlled by Yukawa interactions, plays an equally important role. To illustrate its impact, we  turn to a brief discussion of  weakly coupled gauge theories
from the viewpoint of their phase diagrams.
Four distinct cases arise: Besides the Gaussian fixed point, gauge theories either display  none, the Banks-Zaks, gauge-Yukawa, or the Banks-Zaks and gauge-Yukawa fixed points, depending on the values for $B, C$, and $C'$, see  Tab.~\ref{tFP}~{\bf c)}. The different phase diagrams are shown qualitatively in Figs.~\ref{pAFa} --\ref{pAFd},  projected onto  the $(\alpha,Y_4)$ plane.   
\step

Gauge theories with  $B>0$ and $C<0$ have no weakly coupled fixed points.
At weak coupling,  the phase diagram solely displays asymptotic freedom
and the Gaussian UV fixed point, Fig.~\ref{pAFa}. The  set of UV free trajectories emanating out of it are indicated by the red shaded area.
 Its upper boundary is provided by the Yukawa nullcline which also acts as an infrared attractor  \cite{Ghika:1975zv,Maiani:1977cg,Froggatt:1978nt,Iliopoulos:1980zd,Pendleton:1980as}  
due to the fact that the sign of \eq{BetaYukawa} is always controlled by the gauge field fluctuations for small Yukawa couplings.  
On the scaling trajectory, the gauge, Yukawa and scalar couplings run at the same rate into the Gaussian UV fixed point \cite{Chang:1974bv}. UV free trajectories continue towards the domain of strong coupling where the theory is expected to display confinement and chiral symmetry breaking, or, possibly, a strongly coupled IR fixed point. On the other hand, above the Yukawa nullcline no trajectories are found which can reach the Gaussian in the UV. On such trajectories, the theory technically loses asymptotic freedom. Predictivity is then limited up to a finite UV scale, unless a strongly coupled UV fixed point materialises out of the blue.  
\step

Gauge theories with $B>0$ and $C>0>C'$  additionally develop a  Banks-Zaks fixed point \eq{BC} which is perturbative provided  $B/C$ is sufficiently small. Yukawa couplings are immaterial for this. Banks-Zaks fixed points are always weakly attractive in the gauge and strongly repulsive in the Yukawa direction. The former follows from asymptotic freedom together with \eq{Yukawa}, while the latter follows from  \eq{BetaYukawa} and
 $\partial \bm F^A/\partial\bm Y^B$ being non-negative and proportional to the gauge coupling times the  sum of the quadratic Casimirs of the fermions attached to the vertex. 
Moreover, at weak coupling and close to the Banks-Zaks, the flow is always parametrically faster into the $Y_4$ than into the gauge direction. Consequently, the Bank-Zaks fixed point together with the Yukawa nullcline act as a strong infrared-attractive funnel for all trajectories emanating 
from the Gaussian UV fixed point, see Fig.~\ref{pAFb}.  This leads
to low energy relations between the Yukawa and the gauge coupling dictated by \eq{BetaYukawa} (at weak coupling), irrespective of their detailed UV origin.\footnote{Exact examples are given by the gauge-Yukawa theories of \cite{Litim:2014uca} in the parameter range $0<11/2-N_F/N_C\ll 1$.} 
Elsewise the same discussion as in the previous example applies.\step

Progressing towards gauge theories with $B>0$ and $C>C'>0$ we now additionally observe a fully interacting gauge-Yukawa fixed point besides the Banks-Zaks, displayed in Fig.~\ref{pAFc}.
The main new effect in theories with $C'>0$ as opposed to those with $C'<0$  is that the funneling of flow trajectories towards the IR attractive Yukawa nullcline comes to a halt, whereby couplings take an interacting IR  fixed point \eq{YukawaFP}, \eq{BC'}. Furthermore, the fixed point is genuinely  attractive in both the gauge and the Yukawa directions.\footnote{In theories with several Yukawa couplings several gauge-Yukawa fixed point may arise of which at least one is fully IR attractive. See \cite{Terao:2007jm} for an explicit example with a single Yukawa coupling.} 
The theory comes out more strongly coupled at the gauge-Yukawa   than at the Banks-Zaks fixed point owing to \eq{lower}. 
The gauge-Yukawa fixed point  characterises a second order phase transition between a symmetric phase and a phase with spontaneous symmetry breaking where the scalars acquire a non-vanishing vacuum expectation value. Details of the phase transition becomes visible once  mass terms are added, taking the role of temperature, with the scalar vacuum expectation values serving as order parameters. Spontaneous symmetry breaking may also entail the breaking of chiral symmetry via Yukawa couplings. Away from fixed points, the theory may display a number of further phenomena such as first order phase transitions, dimensional transmutation, decoupling, and confinement in the deep IR.\footnote{Phenomenological aspects of IR gauge-Yukawa fixed points have been pioneered in \cite{Terao:2007jm,Grinstein:2011dq} (see also \cite{Antipin:2011aa,Antipin:2012sm}).   Models with gauge-Yukawa fixed points have also been studied from the viewpoint of conformal field theory \cite{Luty:2012ww} and the $a$~theorem \cite{Antipin:2013pya}.}  
\step

Turning to simple or abelian gauge theories with  $B<0$ and $C'<0$ we observe that asymptotic freedom is absent and the Gaussian has become an infrared fixed point. Also, it is impossible for this type of theories to have a Banks-Zaks fixed point owing to the no go theorem \eq{nogo}. However, the Yukawa interactions have turned the two loop coefficient $C>0$ effectively into  $C'<0$ allowing for  an  interacting gauge-Yukawa fixed point \eq{BC'} as displayed in Fig.~\ref{pAFd}.
This fixed point genuinely displays an attractive and a repulsive direction, the former being a consequence of the IR attractive nature of Yukawa nullclines, and the  latter a consequence of infrared freedom in the gauge coupling. Moreover, it qualifies as an asymptotically safe fixed point owing to the two UV finite trajectories emanating out of it \cite{Litim:2014uca}.  
The weak coupling trajectory connects the interacting fixed point with the Gaussian in the infrared whereby the theory remains unconfined at all scales. The strong coupling trajectory, as in the previous cases, is expected to lead to confinement and chiral symmetry breaking, or conformal behaviour at low energies. Away from the Yukawa nullcline (which always coincides with the hypercritical surface of the gauge-Yukawa fixed point), no trajectories are found which can reach the gauge-Yukawa fixed point in the UV. 
On such trajectories,  the theory technically loses asymptotic safety and predictivity is limited by a maximal UV scale unless a novel  UV fixed point emerges at strong coupling.
\step

As an aside, it is worth noticing a  similarity between  gauge-Yukawa theories with complete asymptotic freedom and a Banks-Zaks, and gauge-Yukawa  theories with  asymptotic safety, see Figs.~\ref{pAFb} and~\ref{pAFd}. In both cases, trajectories which escape from the UV fixed point region  towards strong coupling in the IR are solely determined by the Yukawa nullcline. All settings predict IR relations between Yukawa and gauge couplings.
In the former case this arises due to a funnel effect while in the latter it follows from the unstable direction of the  interacting UV fixed point. Without Banks-Zaks, IR relations may be avoided at the expense of substantial fine-tuning in the deep UV, see Fig.~\ref{pAFa}.
 \step

The discussion of phase diagrams generalises  to more complex settings. Gauge theories with several independent Yukawa  couplings will lead to several parameters $C'$, which, depending on their magnitudes, may generate several gauge-Yukawa fixed points.  
Phase diagrams  will then display an enhanced structure owing to additional cross-over phenomena amongst the various fixed points. An even richer pattern arises for
theories with product gauge groups, see Tab.~\ref{tFP}~{\bf d)}. Here, the gauge loop coefficients $B_a$ and  $C_{ab}$ together with the Yukawa-induced coefficients $B'_a$ 
uniquely determine the fixed point structure at weak coupling. 
Evidently, for each gauge coupling individually our discussion based on the ``diagonal'' coefficients $B$, $C$ and $C'$ applies, meaning that parts of the  enlarged phase diagrams materialise as ``direct products'' of those shown in Figs.~\ref{pAFa}--\ref{pAFd}. As a novel addition, theories  will also display ``off-diagonal'' Banks-Zaks and gauge-Yukawa fixed points as well as fully interacting products thereof, depending on the availability and structure of the solutions to \eq{FPa'b}.\footnote{See \cite{Esbensen:2015cjw} for a recent example  in semi-simple gauge theories without Yukawa couplings.} 
Furthermore, each interacting fixed point naturally relates to a conformal window similar to those of QCD with fermionic matter.
Some of the fixed points of (product) gauge theories offer UV conformal windows around fixed points with exact asymptotic safety at weak coupling. It is therefore natural to speculate that some such models may qualify as UV completions for the Standard Model of particle physics. \step
  
{\bf 11.}  Finally, we  briefly comment on interacting fixed points in $4d$ supersymmetric QFTs. Supersymmetry imposes relations amongst gauge, Yukawa, and scalar couplings \cite{Weinberg:2000cr}.  In general, quartic scalar selfinteractions 
are no longer independent. For theories with $N=1$ supersymmetry without superpotentials,
gauge beta functions remain of the form \eq{eq:GaugeBeta} at weak coupling. The signs of $B$ and $C$ depend on the matter content \cite{Jones:1974pg}. Gauge sectors can develop  Banks-Zaks fixed points \eq{BC} which are always IR $(B>0)$ but never UV \cite{Martin:2000cr}, fully consistent with our findings in non-supersymmetric theories \eq{nogo}, \eq{nogoa}. An important difference arises once superpotentials (i.e.~Yukawa couplings) are present.  Owing to supersymmetry, Yukawas can only take weakly interacting fixed points provided at least one of the gauge sectors is asymptotically free  \cite{Martin:2000cr}. This implies that asymptotic safety at weak coupling is out of reach for simple $N=1$ supersymmetric gauge theories. Overall, weakly interacting fixed points are either absent, or of the Banks-Zaks, or of the gauge-Yukawa~type. Phase diagrams of simple $4d$ gauge theories with $N=1$ supersymmetry take the form Fig.~\ref{pAFa} or Fig.~\ref{pAFc}, while settings with Fig.~\ref{pAFb} or Fig.~\ref{pAFd} cannot be realised.
 For $N=2$ supersymmetry, Yukawa couplings are no longer independent but related to the gauge coupling. Moreover, the running of the gauge coupling becomes one-loop exact with \eq{eq:GaugeBeta} and $C\equiv 0$ \cite{Howe:1983wj,Jones:1983vk}. Hence, $N=2$ theories are either asymptotically free or infrared free and interacting fixed points cannot arise. In the limit where $B=0$,  the gauge coupling becomes exactly marginal leading to a line of fixed points \cite{Howe:1983wj}. The latter continues to hold true for maximally extended supersymmetry, $N=4$ SYM, where the  constraints from supersymmetry are so powerful that the theory does not flow under the RG, and  any value of the gauge coupling corresponds to a fixed point.\footnote{For further constraints on supersymmetric fixed points including at strong coupling, see \cite{Martin:2000cr,Intriligator:2015xxa}.}

   {\bf 12.} 
  In summary, we have identified the  interacting  fixed points  of  four-dimensional  gauge theories  in the regime where gauge and matter fields remain good fundamental degrees of freedom.  Low-energy fixed points are either of the Banks-Zaks or gauge-Yukawa type, or combinations and products theoreof (Tab.~\ref{tFP}), offering a rich spectrum of phenomena including phase transitions and the spontaneous breaking of symmetry. 
We have  also derived  no go theorems together with necessary and sufficient conditions to guarantee asymptotic safety  of general gauge theories  (Tab.~\ref{tFinal}). 
Interacting high-energy fixed points are invariably of the gauge-Yukawa type and require elementary scalar fields such as the Higgs. Hence, the findings of \cite{Litim:2014uca} were not a coincidence: rather,   the dynamical mechanism to tame the notorious Landau poles 
of general infrared free gauge theories is {\it unique}, and, owing to the group-theoretical limitation \eq{bound}, {\it exclusively} delivered through
Yukawa interactions.  We conclude that our findings open a window of opportunities towards perturbative UV completions of the Standard Model beyond the paradigm of asymptotic freedom. \step

\step
\step
\step

\centerline{\bf Acknowledgements} 

We thank Tom Banks, Steve Carlip, Gian Giudice, David Gross, Petr Horava, Hugh Osborn, Joe Polchinski, Graham Ross, Martin Schmaltz, Witold Skiba, and Gabriele Veneziano for comments and discussions. 
Some  of the results have been presented at  the workshops {\it Particle Phenomenology from the Early Universe to High Energy Colliders} (Portoroz, Apr 2015), {\it The Origin of Mass} (Odense, May 2015), {\it Quantum Gravity Foundations: UV to IR} (KITP, Santa Barbara, Jun 2015), {\it Probing the Mystery: Theory and Experiment in Quantum Gravity} (Galiano, Jul 2015), and {\it Lattice QCD in the Era of the LHC} (KITP, Santa Barbara, Aug 2015). DFL thanks the organisers for their hospitality and financial support. 
This work is supported by the Science and Technology Research Council under the Consolidated Grant [ST/G000573/1] and by an STFC Studentship.

\bibliographystyle{apsrev4-1}
\bibliography{NoGobib}

 \end{document}